# Vortex Breit-Wheeler Electron-positron pair creation via vortex gamma-photons

Zhigang Bu,[1] Liangliang Ji,[1,*] Shaohu Lei,[1,2] Huayu Hu,[3] Xiaomei Zhang,[4] and Baifei Shen[4,1,†]

[1]*State Key Laboratory of High Field Laser Physics and CAS Center for Excellence in Ultra-intense Laser Science, Shanghai Institute of Optics and Fine Mechanics (SIOM), Chinese Academy of Sciences (CAS), Shanghai 201800, China.*

[2] *University of Chinese Academy of Sciences, Beijing 100049, China*

[3]*Hypervelocity Aerodynamics Institute, China Aerodynamics Research and Development Center, Mianyang, Sichuan 621000, China*

[4]*Department of Physics, Shanghai Normal University, Shanghai, 200234, China*

**Abstract:** Particles in quantum vortex states (QVS) carrying definite orbital angular momenta (OAM) brings new perspectives in various fundamental interaction processes. When unique properties arise in the QVS, understanding how OAM manifest itself between initial particles and the outcome in vortex particle collisions becomes essential. This is made possible by applying the complete vortex description for all involved particles such that angular momenta (AM) are represented by explicit quantum numbers and their connections are naturally retrieved. We demonstrate the full-vortex quantum-electrodynamics (QED) results for the Breit-Wheeler pair creation process and derive the AM-dependent selection rule. The numerically resolved cross-sections show anti-symmetric spin polarization and most importantly, the first OAM spectra in vortex collision processes. The latter reveals efficient conversion of OAM to created pairs, leading to featured hollow and ring-shaped structure in the density distribution. These results demonstrate a clear picture in understanding the OAM physics in the scattering processes of high energy particles.

E-mail: * jill@siom.ac.cn; † bfshen@mail.shcnc.ac.cn

# 1 Introduction

The experimental advances in production and measurement of electrons carrying orbital angular momentum (OAM) [1-4] have attracted extensive interest in this decade. In addition to spin angular momentum (SAM), electron beams containing OAM are closely associated with the vortex structure, also known as the "twisted" modes. Theoretical description of the relativistic vortex electrons is based on the spinor Bessel mode [5,6] as an exact solution of the Dirac equation. In analog to photon [7] and scalar particle states [8], the quantum vortex state (QVS) in the Bessel mode can be interpreted as the superposition of the plane wave states according to particular phase structure. Alternatively, direct comparison with the optical vortex field [9,10] leads to another vortex state, the Laguerre-Gauss mode [11,12], under the paraxial approximation. These vortex states have shown several unique features, including spin-orbital coupling [5,13], spin-orbital conversion [14], spin and orbital Hall effects [15,16] and geometric phase [17]. The OAM degree of freedom manifest itself in various interaction processes such as vortex electron state dressed in laser field [18,19], Compton scattering [7], elastic electron-atom scattering [20,21], two-photon annihilation of vortex positron [22], resonance production in vortex particle collisions [23] and radiation from vortex electrons [24,25]. The on-going studies have brought up new perspectives in high energy physics, nuclear physics, atomic dynamics, strong-field physics and in producing exotic vortex light sources.

In a broad range of particle collisions, transferring of polarization and OAM from the initial states to the final ones is of central interest. Governed by the conservation law of total angular momentum (TAM), this important connection can be revealed from the angular momentum (AM) dependent cross-sections. The latter depends on resolving the complete vortex scattering in which all interacting particles are described by the QVS, as a plane wave does not carry OAM. Up to now, however, due to the complexity of the QVS, the quantum electrodynamics (QED) picture of complete vortex scattering remains to be seen.

In this article, we apply the first full-vortex description on the Breit-Wheeler (BW) pair creation process, where OAM properties appear in both photon fields and created fermion particles. To describe the annihilation of two QVS gamma-photons into a QVS electron-positron pair, we employ the vortex states for both and derive the scattering cross section within the theoretical framework of QED. This allows AM to be explicitly expressed by the quantum numbers and inherently conserved. The AM-dependent selection rules and the OAM spectra are illustrated for the first time. Together they show how OAM is transferred from the initial gamma photons to final electrons and positrons, which also depends on the asymmetric polarization distribution. Consequently, the generated pairs exhibit periodic ring-shape and hollow structure in the density distribution, providing a unique feature for identifying the vortex scattering process.

This article is organized as follows. In section 2 we briefly review the strict definition of QVS, which is the basis of performing the field quantization and calculating the S-matrix of a vortex scattering process. In section 3 we derive the S-matrix and the cross-section of the vortex BW pair creation process, in which the

AM-dependent selection rule is obtained. In section 4 we discuss the numerical results, including the momentum and AM spectra, the polarization characteristic of the scattering, the density distribution of the created pair is also discussed. In section 5 we conclude.

## 2 Definition of quantum vortex states: Bessel modes

The particles carrying definite OAM can be described by QVS (also known as "twisted state"). The QVS is defined by the Bessel modes which are constructed from the superposition of plane wave (PW) states via $\Phi^{s,l}_{k_\perp,k_z}(x) = \int d^3 k' \tilde{\varphi}^l_{k_\perp,k_z}(k') \Phi^s_{k'}(x)$ in cylindrical momentum space $k' = (k'_\perp, \phi'_k, k'_z)$ [7], assuming the particle propagates along z-axis. Here $\Phi^s_{k'}(x)$ denotes a PW scalar, vector or spinor particle state, and $\tilde{\varphi}^l_{k_\perp,k_z}(k') = 1/(\sqrt{2\pi} i^l k_\perp) \delta(k'_z - k_z) \delta(k'_\perp - k_\perp) e^{il\phi'_k}$ is the Fourier spectrum containing a vortical phase with OAM number $l$, $s$ represents the spin if the particle carries. Under this definition, the QVS of photon field takes the form

$$A^{j,\lambda;\mu}_{k_\perp,k_z}(x) = \varepsilon^{j,\lambda;\mu}_{k_\perp,k_z}(r) e^{ik_z z - i\omega t}$$

$$= \frac{e^{ik_z z - i\omega t}}{4\pi\sqrt{\omega}} \begin{pmatrix} 0 \\ (i/2)\left[(1-k_z/\omega)\Theta^{j+\lambda}_{k_\perp}(r) + (1+k_z/\omega)\Theta^{j-\lambda}_{k_\perp}(r)\right] \\ (\lambda/2)\left[(1-k_z/\omega)\Theta^{j+\lambda}_{k_\perp}(r) - (1+k_z/\omega)\Theta^{j-\lambda}_{k_\perp}(r)\right] \\ (\lambda k_\perp/\omega)\Theta^{j}_{k_\perp}(r) \end{pmatrix}, \quad (1)$$

where $\lambda=\pm 1$ is the polarization parameter, $\omega$ is the photon energy, $j$ is the total angular momentum (TAM) of the QVS photon and the transverse function is defined by $\Theta^n_{k_\perp}(r) = J_n(k_\perp r) e^{in\theta}$ ($J_n(r)$ is the Bessel function of the first kind). The photon QVS satisfies the orthonormality: $\left(A^{j,\lambda}_{k_\perp,k_z}, A^{j',\lambda'}_{k'_\perp,k'_z}\right) = -(1/k_\perp)\delta_{\lambda\lambda'}\delta_{jj'}\delta(k_\perp - k'_\perp)\delta(k_z - k'_z)$. We use the natural units $\hbar=c=1$ throughout the derivation. Analogously, the QVS of electron and positron can be constructed from the positive and negative-frequency

PW solutions of the Dirac equation, and they are given by,

$$\psi^{\pm,l,s}_{p_\perp,p_z}(x) = \frac{e^{\pm ip_z z \mp iEt}}{\sqrt{2}(2\pi)|p|}\sqrt{1-\frac{M}{E}}\left[\chi^\pm_z \Theta^l_{p_\perp}(r) \pm ip_\perp \chi^\pm_\perp\right], \tag{2}$$

where $M$ and $E$ are electron/positron mass and energy. $\chi^\pm_z$ and $\chi^\pm_\perp$ represent the spinor wave functions in longitudinal and transverse dimensions. For electron, $\chi^+_z = \begin{pmatrix}(E+M)\xi^s \\ p_z\sigma_z\xi^s\end{pmatrix}$ and $\chi^+_\perp = \begin{pmatrix} 0 \\ \sigma^{l,p_\perp}_\perp \xi^s\end{pmatrix}$; for positron, $\chi^-_z = \begin{pmatrix} p_z\sigma_z\eta^s \\ (E+M)\eta^s\end{pmatrix}$ and $\chi^-_\perp = \begin{pmatrix}\sigma^{l,p_\perp}_\perp \eta^s \\ 0\end{pmatrix}$, with transversal matrix $\sigma^{l,p_\perp}_\perp = \begin{pmatrix} 0 & -\Theta^{l-1}_{p_\perp}(r) \\ \Theta^{l+1}_{p_\perp}(r) & 0 \end{pmatrix}$. The two-component spinors $\xi^s$ and $\eta^s$ characterize the electron and positron spins in the rest frame. The spinor QVS satisfies the orthonormality: $\left(\psi^{\pm,l,s}_{p_\perp,p_z},\psi^{\pm,l',s'}_{p'_\perp,p'_z}\right) = (1/p_\perp)\delta_{ss'}\delta_{ll'}\delta(p_\perp - p'_\perp)\delta(p_z - p'_z)$. One finds that due to the spin-orbit interaction of relativistic particles, the QVS is neither the eigenmodes of the OAM operator $\hat{L}_z$, nor the SAM operator $\hat{S}_z$ along the z-direction, but rather of the TAM operator $\hat{J}_z = \hat{L}_z + \hat{S}_z$. For a vortex photon, the eigenvalue of $\hat{J}_z$ is $j$: $\hat{J}_z A^{j,\lambda}_{k_\perp,k_z,\mu} = jA^{j,\lambda}_{k_\perp,k_z,\mu}$; the vortex electron/positron gives $\hat{J}_z \psi^{\pm,l,s}_{p_\perp,p_z} = (l+s/2)\psi^{\pm,l,s}_{p_\perp,p_z}$.

**3 S-matrix and scattering cross section of vortex BW process**

In perturbation theory, the S-matrix for the vortex BW process contains two terms: $S_{fi} = S_1 + S_2$. The Feynman diagram for $S_1$ is drawn in Fig. 1(a), where the cone on each external line represents a QVS and its polar angle is defined by $\tan\alpha_p = p_\perp/p_z$. This diagram is calculated as

$$S_1 = -ie^2\int d^4x d^4x' \frac{d^4q}{(2\pi)^4} \overline{\psi}^{+,l_1,s_1}_{p_{1\perp},p_{1z}}(x) \slashed{A}^{j_1,\lambda_1}_{k_{1\perp},k_{1z}}(x) \frac{(\slashed{q}+M)}{(q^2-M^2)} e^{-iq\cdot(x-x')} \slashed{A}^{j_2,\lambda_2}_{k_{2\perp},k_{2z}}(x')\psi^{-,l_2,s_2}_{p_{2\perp},p_{2z}}(x') \tag{3}$$

with the slash $\slashed{A} = A_\mu\gamma^\mu$. We consider the forward and backward scatterings here. Substituting the eigenmodes of photon field (1) and electron/positron field (2) into Eq.

(3), one obtains

$$S_1 = -\frac{ie^2}{16\pi}\sqrt{\frac{(E_1-M)(E_2-M)}{\omega_1\omega_2 E_1 E_2 \boldsymbol{p}_1^2 \boldsymbol{p}_2^2}}\delta(\omega_1+\omega_2-E_1-E_2)\delta(k_{1z}+k_{2z}-p_{1z}-p_{2z}),$$
$$\times \xi^{s_1\dagger}\Xi^{j_1,\lambda_1;j_2,\lambda_2}_{k_{1\perp},k_{1z};k_{2\perp},k_{2z}}(l_1,p_{1\perp},p_{1z};l_2,p_{2\perp},p_{2z})\eta^{s_2}\Big|_{\substack{E_q=-\omega_1+E_1=\omega_2-E_2\\q_z=-k_{1z}+p_{1z}=k_{2z}-p_{2z}}} \quad (4)$$

where the $(2\times 2)$ matrix

$$\Xi^{j_1,\lambda_1;j_2,\lambda_2}_{k_{1\perp},k_{1z};k_{2\perp},k_{2z}}(l_1,p_{1\perp},p_{1z};l_2,p_{2\perp},p_{2z}) = \begin{pmatrix} \upsilon_{11}\delta_{j_1+j_2,l_1-l_2} & \upsilon_{12}\delta_{j_1+j_2,l_1-l_2+1} \\ \upsilon_{21}\delta_{j_1+j_2,l_1-l_2-1} & \upsilon_{22}\delta_{j_1+j_2,l_1-l_2} \end{pmatrix}. \quad (5)$$

The matrix elements $\upsilon_{11}$, $\upsilon_{12}$, $\upsilon_{21}$ and $\upsilon_{22}$ are derived by Eq. (A10)-(A13) in appendix A. The second term $S_2$ can be obtained by exchanging two incident photons in $S_1$, $\Xi$ matrix and matrix elements $\upsilon$: $S_2,\tilde{\Xi},\tilde{\upsilon} = S_1,\Xi,\upsilon(k_1\leftrightarrow k_2, j_1\leftrightarrow j_2, \lambda_1\leftrightarrow \lambda_2)$. It is found that each matrix element in $\Xi$ and $\tilde{\Xi}$ has an AM-dependent Kronecker delta function, which gives the selection rules for the vortex BW process,

$$j_1+j_2 = l_1-l_2+\Delta, \quad (6)$$

with $\Delta=0, \pm 1$. The minus sign before $l_2$ stems from the definition of positron AM. Using the S-matrix element one obtains the spin-dependent pair creation probability: $d\mathcal{P} = (2\pi^2/(RL))^2 k_{1\perp}k_{2\perp}p_{1\perp}p_{2\perp}|S_{fi}|^2 dp_{1\perp}dp_{1z}dp_{2\perp}dp_{2z}$, where $R$ and $L$ are the large radius and length used in the cylindrical normalization. The differential cross section is derived as $d\sigma = d\mathcal{P}/(T\langle u_z\rangle)$. For head-on collision, the relative current density of the incident photons is given by $\langle u_z\rangle = |k_{1z}/\omega_1 - k_{2z}/\omega_2|/V$, and the cross section reads

$$d\sigma = \frac{\pi^3\alpha^2}{16}\frac{k_{1\perp}k_{2\perp}p_{1\perp}p_{2\perp}(E_1-M)(E_2-M)}{\omega_1\omega_2 E_1 E_2 \boldsymbol{p}_1^2 \boldsymbol{p}_2^2 |k_{1z}/\omega_1-k_{2z}/\omega_2|}\mathrm{Tr}\left[\xi^{s_1}\xi^{s_1\dagger}(\Xi+\tilde{\Xi})\eta^{s_2}\eta^{s_2\dagger}(\Xi+\tilde{\Xi})^\dagger\right]$$
$$\times \delta(\omega_1+\omega_2-E_1-E_2)\delta(k_{1z}+k_{2z}-p_{1z}-p_{2z})dp_{1\perp}dp_{1z}dp_{2\perp}dp_{2z} \quad (7)$$

For the unpolarized photon scattering, the cross section can be obtained by averaging over the photon polarization $\lambda_1$ and $\lambda_2$.

The selection rules from Eq. (6) are related to the spin states of electron and position in the rest frame. If $\xi^{s_1}$ is the eigenmode of $\sigma_z$ for electron with the eigenvalue $s_1=+1$: $\xi^{s_1} = \begin{pmatrix} 1 \\ 0 \end{pmatrix}$, and $\eta^{s_2}$ for positron with $s_2=+1$: $\eta^{s_2} = \begin{pmatrix} 1 \\ 0 \end{pmatrix}$, the trace in Eq. (7) is given by $\text{Tr}\left[\xi^{s_1}\xi^{s_1\dagger}\left(\Xi+\tilde{\Xi}\right)\eta^{s_2}\eta^{s_2\dagger}\left(\Xi+\tilde{\Xi}\right)^{\dagger}\right] = |\upsilon_{11}+\tilde{\upsilon}_{11}|^2 \delta_{j_1+j_2,l_1-l_2}$. Since positron has the reversed AM definition with respect to electron, i.e., the eigenvalue of the state $\eta^{s_2} = \begin{pmatrix} 1 \\ 0 \end{pmatrix}$ is $m_2=-1/2$, the TAM of the created electron-positron pair is $l_1-l_2$, corresponding to the conservation law: $j_1+j_2=l_1-l_2$, which agrees with the selection rule of $\Delta=0$. Similarly, $\Delta=+1, -1$ correspond to $m_1=m_2=+1/2$ and $m_1=m_2=-1/2$, respectively. The $m_1=-m_2=-1/2$ case is consistent with the selection rule of $\Delta=0$. In other words, Eq. (6) conclude TAM conservation in the vortex BW process. The cross section of non-spin polarized pairs is obtained by summing over the pair spin states.

**4 Numerical results**

The cross section for electrons can be numerically obtained by integrating over the positron momentum in Eq. (7). This is done trivially by dropping the two delta functions. We choose an example of two vortex photons with energies $\omega_1=\omega_2=5\text{MeV}$ and z-momenta $k_{1z}=-k_{2z}=4\text{MeV}$. The TAMs of two QVS photons are $j_1=3$ and $j_2=2$, with polarization $\lambda_1=\lambda_2=1$, respectively. Figs. 1(c) and (d) show the differential cross sections of electron and positron, where the AMs are $l_1=3$, $m_1=1/2$ (electron), and $l_2=-2$, $m_2=-1/2$ (positron) following the selection rule $j_1+j_2=l_1-l_2$. We find the cross sections exhibit oscillating distributions in the momentum domain. The structures of electron and positron are symmetrically correlated, governed by the conservation law of energy and momentum. We thus focus on the electron spectrum in the following.

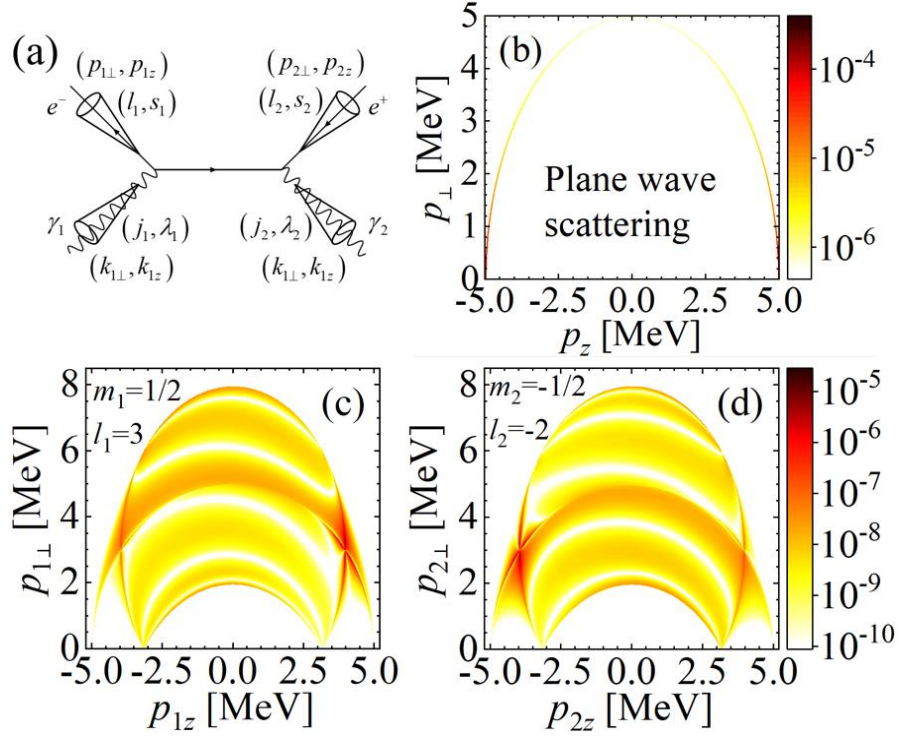

Figure 1. (a) Feynman diagram of matrix element $S_1$, the cones represent the QVS particle. (b) Differential cross sections of electrons, $d\sigma(e^-)/dp_\perp dp_z$[MeV$^{-4}$], in cylindrical momentum space created in BW process with plane-wave particle states. The energies of head-on colliding photons are $\omega_1=\omega_2=5$MeV. Differential cross sections of electrons $d\sigma(e^-)/dp_{1\perp}dp_{1z}$[MeV$^{-4}$] (c) and positrons $d\sigma(e^+)/dp_{2\perp}dp_{2z}$[MeV$^{-4}$] (d) in vortex BW process. The energies and $z$-momenta of the QVS photons are $\omega_1=\omega_2=5$MeV and $k_{1z}=-k_{2z}=4$MeV, the AM parameters are $j_1=3, j_2=2$, $\lambda_1=\lambda_2=1$. The AM of created electron and positron are $l_1=3, m_1=1/2$ and $l_2=-2$, $m_2=-1/2$.

In general, the BW process with PW particle states leads to monoenergetic spectra due to the constraint of four-dimensional momentum conservation. This is shown in Fig. 1(b) as a comparison. The electron cross sections in a head-on collision of two PW photons ($\omega_1=\omega_2=5$MeV) exhibit a half circle with exact 5MeV radius. This is significantly different from the vortex situation (Fig. 1(c) and 1(d)), since the QVS particle has transverse momentum perpendicular to its properation direction. The transverse momentum is characterized by the magnitude, while the direction is destructed by the superposition when constructing the QVS [5,7].

From Fig. 1 one notices that if the QVS gamma-photons are polarized, e.g., $\lambda_1=\lambda_2=1$, the spectral distributions with definite spins are asymmetrical with respect to $\pm|p_z|$, which implies spin polarization for the created particles. The differential cross section $d\sigma(e^-)/d\alpha_{p_1}$ is displayed in Fig. 2. Here we follow the photon energy and momentum employed in Fig. 1 but vary the photon polarization ($\lambda_1$ and $\lambda_2$) and electron/positron spin state ($m_1$ and $m_2$). First of all, we see that unlike the PW BW process which has highest cross-section near the axis (black-solid in Fig. 2(a)), the one for vortex BW process peaks off-axis, due to the transverse momentum in the QVS. Electron with $l_1=3$ and $m_1=+1/2$ (red-solid in Fig. 2(a)) comes from two positron AM channels: $l_2=-5$, $m_2=-1/2$ and $l_2=-4$, $m_2=+1/2$. The channels are switched to $l_2=-6$, $m_2=-1/2$ and $l_2=-5$, $m_2=+1/2$ when the spin is flipped $m_1=-1/2$ (red-dashed). However, these channels are not equally distributed. The former is almost one order of magnitude higher than the latter in the region $\alpha_{p_1} < \pi/2$ and lower in $\alpha_{p_1} > \pi/2$. As a result, electrons emitted into $\alpha_{p_1} < \pi/2$ is characterized with a significant positive polarization ($m_1=+1/2$) and a negative one in opposite directions. This trend also applies for $\lambda_1=\lambda_2=-1$, except that the polarization distributions is reversed, shown in Fig. 2(b).

When photons have opposite polarizations $\lambda_1=-\lambda_2=1$, the cross section with $m_1=+1/2$ is much higher than that with -1/2 in the full angular space $\alpha_{p_1} \in [0,\pi]$, giving +1/2 spin-polarized electrons in all directions as shown in Fig. 2(c). Naturally, $\lambda_1=-\lambda_2=-1$ creates -1/2 spin-polarization (Fig. 2(d)). The electron polarizability along ±z directions is quantified via $\rho = \left(\sigma_{+1/2}(e^-) - \sigma_{-1/2}(e^-)\right) / \left(\sigma_{+1/2}(e^-) + \sigma_{-1/2}(e^-)\right)$ in Fig. 2(e). One sees that $|\rho|$ is above 0.5 at all OAM numbers $l_1$, approaching 0.9 in certain region. Note that for initially unpolarized photons the created electron-positron pair carrying different spins have almost identical distribution, i.e., polarization vanishes.

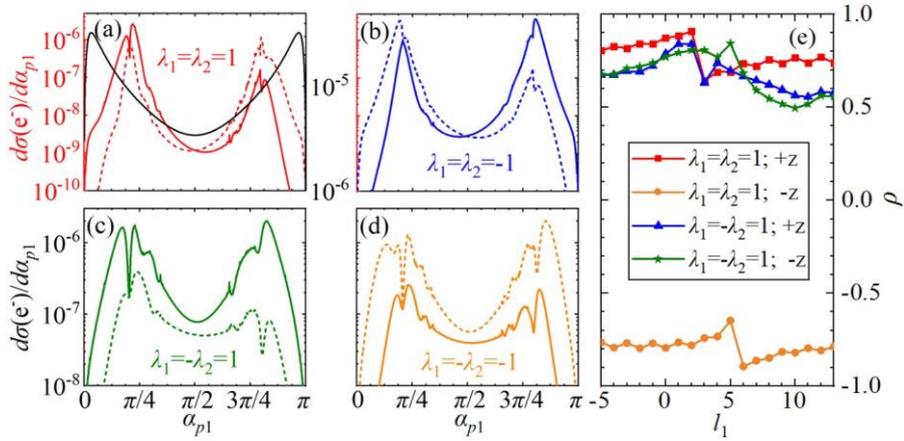

Figure 2. (a)-(d): $\alpha_{p1}$-dependent cross sections $d\sigma(e^-)/d\alpha_{p1}$[MeV$^{-2}$] of electrons carrying the AM of $l_1=3$ for different photon polarizations. The TAMs of the gamma-photons are $j_1=3$ and $j_2=5$, energies $\omega_1=\omega_2=5$MeV and $z$-momenta $k_{1z}=-k_{2z}=4$MeV. The colored solid lines represent the angular distributions of electrons with spin $m_1=1/2$ in the rest frame, the dashed lines represent electrons with spin $m_1=-1/2$. The black solid line in Fig. 2(a) represents the electron angular distribution created in the BW process with plane-wave particle states. (e) Polarizability of electrons scattered into +z and -z directions versus the electron OAM number $l_1$.

A key interest of vortex interaction processes is how it is related to the particle OAM numbers. We show for the first time the OAM-dependent cross-section in Fig. 3. The total cross section peaks at $l_1=2$ and 6, which correspond to the spins $m_1=1/2$ and -1/2, as shown in Fig. 3(a). The peaking values are shifted by one unit with respect to the photon TAM ($j_1=3$, $j_2=5$). In fact, the OAM peak at $l_1=2$ is mainly converted from photon with $j_1=3$ through the dominating channel $m_1=1/2$ and $m_2=-1/2$, see the red lines in Fig. 3(b). The other channel of $m_2=1/2$ is significantly suppressed from the polarization dependence discussed in Fig. 2(a). The blue lines indicate that the peak at $l_1=6$ originates from the photon TAM $j_2=5$, while the polarization channel $m_1=-1/2$ and $m_2=-1/2$ dominates. This connection can also be seen when varying the photon polarization. For instance, with $\lambda_1=-\lambda_2=1$, since the $m_1=1/2$ spin state dominates the electron creation over $m_1=-1/2$, see Fig. 2(c), electrons with OAM number peaking at $l_1=3$ and 6 are suppressed, leading to the main cross section at $l_1=2$ and 5, seen in Fig. 3(c). In other words, the OAM-dependent cross section explicitly reveals the OAM number of photons in certain polarization state. The one-unit shift induced by latter would be negligible at large AM numbers $|j| \gg 1$.

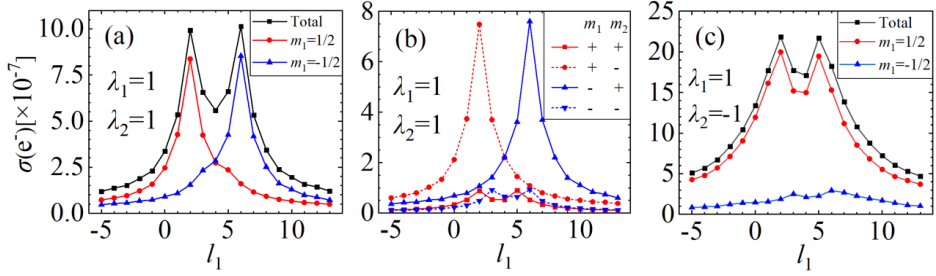

Figure 3. Electron cross sections $\sigma(e^-)[\text{MeV}^{-2}]$ versus electron OAM $l_1$. The incident gamma-photons are polarized, and possess the TAMs $j_1=3$ and $j_2=5$, energies $\omega_1=\omega_2=5\text{MeV}$ and z-momenta $k_{1z}=-k_{2z}=4\text{MeV}$. The photon polarization parameters are $\lambda_1=\lambda_2=1$ (a) and $\lambda_1=-\lambda_2=1$ (c). (b) Electron cross sections $\sigma(e^-)[\text{MeV}^{-2}]$ for $\lambda_1=\lambda_2=1$ polarized photons from different electron/positron polarization channels.

The electron QVS manifests itself in the transverse density distribution. From Eq. (2) we obtain AM-dependent probability density

$$\rho^{+;l_1,s_1}_{\perp;p_{1\perp},p_{1z}} = \frac{1}{2(2\pi)^2}\left[\left(\left(1+\frac{M}{E_1}\right)+\left(1-\frac{M}{E_1}\right)\frac{p_{1z}^2}{|\boldsymbol{p}_1|^2}\right)J_{l_1}^2(p_{1\perp}r)+\frac{p_{1\perp}^2}{|\boldsymbol{p}_1|^2}\left(1-\frac{M}{E_1}\right)J_{l_1+s_1}^2(p_{1\perp}r)\right].$$

(8)

This is shown in Fig. 4(a) of the eigenstate with electron energy $E_1=5\text{MeV}$, transverse momentum $p_{1\perp}=3\text{MeV}$, $l_1$=3 and $m_1$=1/2. A periodic ring-shaped profile is seen with a hollow on the axis. The peak density on each ring declines at larger radii. We determine the radius $R_0$ of the maximum density via $d\rho^{+;l_1,s_1}_{\perp;p_{1\perp},p_{1z}}/dr=0$ (the first ring) and relate it to the polar angle in Fig. 4(b). These dependences suggest a clear connection between the OAM and the radius of the ring structure with known spin states.

In principle, the ring-shaped structure for mono-energetic electrons in Fig. 4(a) could be retrieved by extracting electrons within a small energy spread, implying a potential scheme to identify the AM of high-energy polarized gamma photon. In reality, the created pairs in the vortex BW process are not mono-energetic. When considering the momentum spectrum distribution generated in a scattering process as the weighting factor, the transverse density of the created electrons can be obtained by summing each mono-energetic electron density. From Fig. 4(c) we notice that the transverse density oscillation is significantly weakened or even offset in this case. However, the hollow density structure is always present, as a unique feature of the vortex modes and hence can be seen as a robust signature to identify the vortex scattering process.

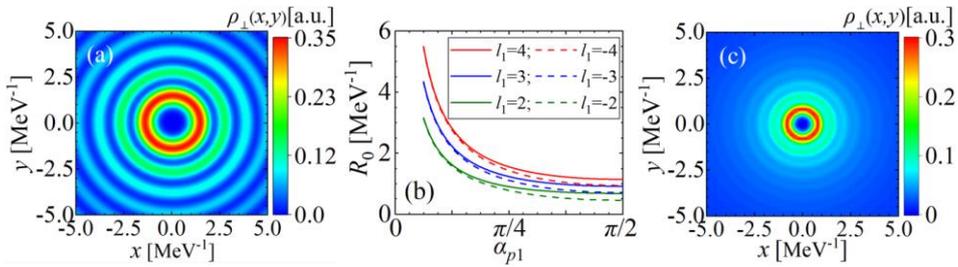

Figure 4. (a) Transverse density distribution of a mono-energetic QVS electron with energy $E_1$=5MeV, transverse momentum $p_{1\perp}$=3MeV, AMs $l_1$=3 and $m_1$=1/2. (b) Radius $R_0$ of first maximum density versus the polar angle of electron momentum for electron spin $s_1=1$ and different OAM $l_1$. (c) Transverse density of the created electrons ($l_1=3$ and $m_1=1/2$) in a vortex BW process.

## 5 Conclusions

In conclusion, we have examined the vortex BW pair creation process based on the QED theory. The full-vortex cross-sections clearly reveal the connections of AM between initial photons and final pairs, showing dependence on polarization and OAM-numbers. The vortex scattering process leads to distinctive hollow and ring-shaped density distribution of the created pairs. These vortex gamma beams can be generated through, for example, high power laser-driven nonlinear Compton Scattering [7,26,27].

## Appendix A: Derivation of S-matrix elements

The $S_1$-matrix in Eq. (3) contains a transverse phase $e^{i q_\perp \cdot r}$ which can be expanded into the AM modes $e^{in\theta}$ in cylindrical coordinates,

$$e^{i q_\perp \cdot r} = \sum_n i^n J_n(q_\perp r) e^{in\theta - in\phi_q}. \tag{A1}$$

Substituting Eq. (1), (2) and (A1) into Eq. (3), one obtains,

$$\begin{aligned}
S_1 = &-\frac{ie^2}{8(2\pi)^3} \sqrt{\frac{(E_1 - M)(E_2 - M)}{\omega_1 \omega_2 E_1 E_2 p_1^2 p_2^2}} \delta(\omega_1 + \omega_2 - E_1 - E_2) \delta(k_{1z} + k_{2z} - p_{1z} - p_{2z}) \\
&\times \sum_{n,n'} i^{n-n'} \int \frac{dq_\perp q_\perp}{(q^2 - M^2)} \Big[\Big(p_{1z} \xi^{s_1 \dagger} \sigma_z \Gamma^{j_1, \lambda_1}_{k_{1\perp}, k_{1z}}, (E_1 + M) \xi^{s_1 \dagger} \Gamma^{j_1, \lambda_1}_{k_{1\perp}, k_{1z}}\Big) \\
&-ip_{1\perp} \Big(\xi^{s_1 \dagger} \tilde{\Gamma}^{j_1, \lambda_1}_{k_{1\perp}, k_{1z}}, 0\Big)\Big] \begin{pmatrix} (E_q + M)\delta_{nn'} & -Q^{n,n'}_{q_\perp, q_z} \\ Q^{n,n'}_{q_\perp, q_z} & -(E_q - M)\delta_{nn'} \end{pmatrix} \\
&\times \Bigg[\begin{pmatrix} (E_2 + M) \Gamma'^{j_2, \lambda_2}_{k_{2\perp}, k_{2z}} \eta^{s_2} \\ -p_{2z} \Gamma'^{j_2, \lambda_2}_{k_{2\perp}, k_{2z}} \sigma_z \eta^{s_2} \end{pmatrix} + ip_{2\perp} \begin{pmatrix} 0 \\ \tilde{\Gamma}'^{j_2, \lambda_2}_{k_{2\perp}, k_{2z}} \eta^{s_2} \end{pmatrix}\Bigg]\Bigg|_{\substack{E_q = -\omega_1 + E_1 = \omega_2 - E_2 \\ q_z = -k_{1z} + p_{1z} = k_{2z} - p_{2z}}},
\end{aligned} \tag{A2}$$

in which there are five (2×2) matrixes:

$$\Gamma^{j_1, \lambda_1}_{k_{1\perp}, k_{1z}} = \int d\theta dr r J_n(q_\perp r) J_{l_1}(p_{1\perp} r) e^{i(n-l_1)\theta} \Lambda^{j_1, \lambda_1}_{k_{1\perp}, k_{1z}}(r, \theta), \tag{A3}$$

$$\tilde{\Gamma}_{k_{1\perp},k_{1z}}^{j_1,\lambda_1} = \int d\theta dr r J_n(q_\perp r) e^{in\theta} \sigma_\perp^{l_1,p_{1\perp}+}(r,\theta) \Lambda_{k_{1\perp},k_{1z}}^{j_1,\lambda_1}(r,\theta), \tag{A4}$$

$$\Gamma'^{j_2,\lambda_2}_{k_{2\perp},k_{2z}} = \int d\theta' dr' r' J_{n'}(q_\perp r') J_{l_2}(p_{2\perp} r') e^{i(l_2-n')\theta'} \Lambda_{k_{2\perp},k_{2z}}^{j_2,\lambda_2}(r',\theta'), \tag{A5}$$

$$\tilde{\Gamma}'^{j_2,\lambda_2}_{k_{2\perp},k_{2z}} = \int d\theta' dr' r' J_{n'}(q_\perp r') e^{-in'\theta'} \Lambda_{k_{2\perp},k_{2z}}^{j_2,\lambda_2}(r',\theta') \sigma_\perp^{l_2,p_{2\perp}}(r',\theta'), \tag{A6}$$

$$Q_{q_\perp,q_z}^{n,n'} = \frac{1}{2\pi} \int d\phi_q e^{-i(n-n')\phi_q} (\boldsymbol{\sigma} \cdot \boldsymbol{q}) = \begin{pmatrix} q_z \delta_{nn'} & q_\perp \delta_{n',n+1} \\ q_\perp \delta_{n',n-1} & -q_z \delta_{nn'} \end{pmatrix}, \tag{A7}$$

with

$$\Lambda_{k_\perp,k_z}^{j,\lambda}(r,\theta) = \begin{pmatrix} \dfrac{\lambda k_\perp}{\omega} J_j(k_\perp r) e^{ij\theta} & i\left(1+\dfrac{\lambda k_z}{\omega}\right) J_{j-1}(k_\perp r) e^{i(j-1)\theta} \\ i\left(1-\dfrac{\lambda k_z}{\omega}\right) J_{j+1}(k_\perp r) e^{i(j+1)\theta} & -\dfrac{\lambda k_\perp}{\omega} J_j(k_\perp r) e^{ij\theta} \end{pmatrix}. \tag{A8}$$

Substituting these equations into (A2), the $S_1$-matrix leads to

$$S_1 = -\frac{ie^2}{16\pi} \sqrt{\frac{(E_1-M)(E_2-M)}{\omega_1\omega_2 E_1 E_2 \boldsymbol{p}_1^2 \boldsymbol{p}_2^2}} \delta(\omega_1+\omega_2-E_1-E_2)\delta(k_{1z}+k_{2z}-p_{1z}-p_{2z}) \\ \times \xi^{s_1\dagger} \begin{pmatrix} \upsilon_{11}\delta_{j_1+j_2,l_1-l_2} & \upsilon_{12}\delta_{j_1+j_2,l_1-l_2+1} \\ \upsilon_{21}\delta_{j_1+j_2,l_1-l_2-1} & \upsilon_{22}\delta_{j_1+j_2,l_1-l_2} \end{pmatrix} \eta^{s_2} \Bigg|_{\substack{E_q=-\omega_1+E_1=\omega_2-E_2 \\ q_z=-k_{1z}+p_{1z}=k_{2z}-p_{2z}}}, \tag{A9}$$

which is consistent with Eq. (4). The four matrix elements in Eq. (A9) are derived as,

$$\upsilon_{11} = \int \frac{dq_\perp q_\perp}{(q^2-M^2)} \Big\{ \big[ p_{1z}p_{2z}q_z + p_{1z}(E_2+M)(E_q+M) + p_{2z}(E_1+M)(E_q-M) \\
+ q_z(E_1+M)(E_2+M) \big] \frac{\lambda_1 k_{1\perp} \lambda_2 k_{2\perp}}{\omega_1 \omega_2} S_{j_1}^{l_1} \tilde{S}_{j_2}^{j_2+l_2} + \big[ p_{1z}p_{2z}q_z - p_{1z}(E_2+M)(E_q+M) \\
- p_{2z}(E_1+M)(E_q-M) + q_z(E_1+M)(E_2+M) \big] \left(1+\frac{\lambda_1 k_{1z}}{\omega_1}\right)\left(1-\frac{\lambda_2 k_{2z}}{\omega_2}\right) S_{j_1-1}^{l_1} \tilde{S}_{j_2+1}^{j_2+l_2+1} \\
+ \big[ p_{1z}p_{2\perp}q_z + p_{2\perp}(E_1+M)(E_q-M) \big] \frac{\lambda_1 k_{1\perp}}{\omega_1}\left(1+\frac{\lambda_2 k_{2z}}{\omega_2}\right) S_{j_1}^{l_1} \tilde{S}_{j_2-1}^{j_2+l_2} \\
+ \big[ p_{1\perp}p_{2z}q_z - p_{1\perp}(E_2+M)(E_q+M) \big] \frac{\lambda_1 k_{1\perp}}{\omega_1}\left(1-\frac{\lambda_2 k_{2z}}{\omega_2}\right) S_{j_1}^{l_1+1} \tilde{S}_{j_2+1}^{j_2+l_2+1} \\
+ \big[ p_{1z}p_{2z}q_\perp + q_\perp(E_1+M)(E_2+M) \big] \frac{\lambda_1 k_{1\perp}}{\omega_1}\left(1-\frac{\lambda_2 k_{2z}}{\omega_2}\right) S_{j_1}^{l_1} \tilde{S}_{j_2+1}^{j_2+l_2+1} \\
- \big[ p_{1z}p_{2z}q_\perp + q_\perp(E_1+M)(E_2+M) \big] \frac{\lambda_2 k_{2\perp}}{\omega_2}\left(1+\frac{\lambda_1 k_{1z}}{\omega_1}\right) S_{j_1-1}^{l_1} \tilde{S}_{j_2}^{j_2+l_2} \\
+ \big[ p_{1z}p_{2\perp}q_z - p_{2\perp}(E_1+M)(E_q-M) \big] \frac{\lambda_2 k_{2\perp}}{\omega_2}\left(1+\frac{\lambda_1 k_{1z}}{\omega_1}\right) S_{j_1-1}^{l_1} \tilde{S}_{j_2}^{j_2+l_2+1}$$

$$+\left[p_{1\perp}p_{2z}q_z+p_{1\perp}(E_2+M)(E_q+M)\right]\frac{\lambda_2 k_{2\perp}}{\omega_2}\left(1-\frac{\lambda_1 k_{1z}}{\omega_1}\right)S_{j_1+1}^{l_1+1}\tilde{S}_{j_2}^{j_2+l_2}$$

$$+p_{1z}p_{2\perp}q_\perp\frac{\lambda_1 k_{1\perp}\lambda_2 k_{2\perp}}{\omega_1\omega_2}S_{j_1}^{l_1}\tilde{S}_{j_2}^{j_2+l_2+1}+p_{1\perp}p_{2\perp}q_z\frac{\lambda_1 k_{1\perp}\lambda_2 k_{2\perp}}{\omega_1\omega_2}S_{j_1}^{l_1+1}\tilde{S}_{j_2}^{j_2+l_2+1}$$

$$-p_{1\perp}p_{2z}q_\perp\frac{\lambda_1 k_{1\perp}\lambda_2 k_{2\perp}}{\omega_1\omega_2}S_{j_1}^{l_1+1}\tilde{S}_{j_2}^{j_2+l_2}-p_{1\perp}p_{2\perp}q_\perp\frac{\lambda_1 k_{1\perp}}{\omega_1}\left(1+\frac{\lambda_2 k_{2z}}{\omega_2}\right)S_{j_1}^{l_1+1}\tilde{S}_{j_2-1}^{j_2+l_2}$$

$$+p_{1\perp}p_{2\perp}q_\perp\frac{\lambda_2 k_{2\perp}}{\omega_2}\left(1-\frac{\lambda_1 k_{1z}}{\omega_1}\right)S_{j_1+1}^{l_1+1}\tilde{S}_{j_2}^{j_2+l_2+1}-p_{1z}p_{2\perp}q_\perp\left(1+\frac{\lambda_1 k_{1z}}{\omega_1}\right)\left(1+\frac{\lambda_2 k_{2z}}{\omega_2}\right)S_{j_1-1}^{l_1}\tilde{S}_{j_2-1}^{j_2+l_2}$$

$$+p_{1\perp}p_{2\perp}q_z\left(1-\frac{\lambda_1 k_{1z}}{\omega_1}\right)\left(1+\frac{\lambda_2 k_{2z}}{\omega_2}\right)S_{j_1+1}^{l_1+1}\tilde{S}_{j_2-1}^{j_2+l_2}$$

$$+p_{1\perp}p_{2z}q_\perp\left(1-\frac{\lambda_1 k_{1z}}{\omega_1}\right)\left(1-\frac{\lambda_2 k_{2z}}{\omega_2}\right)S_{j_1+1}^{l_1+1}\tilde{S}_{j_2+1}^{j_2+l_2+1}\Bigg\},$$

(A10)

$$\upsilon_{12}=-i\int\frac{dq_\perp q_\perp}{(q^2-M^2)}\Bigg\{\left[p_{1z}p_{2z}q_z-p_{1z}(E_2+M)(E_q+M)+p_{2z}(E_1+M)(E_q-M)\right.$$

$$\left.-q_z(E_1+M)(E_2+M)\right]\frac{\lambda_1 k_{1\perp}}{\omega_1}\left(1+\frac{\lambda_2 k_{2z}}{\omega_2}\right)S_{j_1}^{l_1}\tilde{S}_{j_2-1}^{j_2+l_2-1}$$

$$+\left[p_{1z}p_{2z}q_z+p_{1z}(E_2+M)(E_q+M)-p_{2z}(E_1+M)(E_q-M)\right.$$

$$\left.-q_z(E_1+M)(E_2+M)\right]\frac{\lambda_2 k_{2\perp}}{\omega_2}\left(1+\frac{\lambda_1 k_{1z}}{\omega_1}\right)S_{j_1-1}^{l_1}\tilde{S}_{j_2}^{j_2+l_2}$$

$$+\left[p_{1z}p_{2z}q_\perp-q_\perp(E_1+M)(E_2+M)\right]\frac{\lambda_1 k_{1\perp}\lambda_2 k_{2\perp}}{\omega_1\omega_2}S_{j_1}^{l_1}\tilde{S}_{j_2}^{j_2+l_2}$$

$$-\left[p_{1z}p_{2\perp}q_z+p_{2\perp}(E_1+M)(E_q-M)\right]\frac{\lambda_1 k_{1\perp}\lambda_2 k_{2\perp}}{\omega_1\omega_2}S_{j_1}^{l_1}\tilde{S}_{j_2}^{j_2+l_2-1}$$

$$+\left[p_{1\perp}p_{2z}q_z+p_{1\perp}(E_2+M)(E_q+M)\right]\frac{\lambda_1 k_{1\perp}\lambda_2 k_{2\perp}}{\omega_1\omega_2}S_{j_1}^{l_1+1}\tilde{S}_{j_2}^{j_2+l_2}$$

$$-\left[p_{1z}p_{2z}q_\perp-q_\perp(E_1+M)(E_2+M)\right]\left(1+\frac{\lambda_1 k_{1z}}{\omega_1}\right)\left(1+\frac{\lambda_2 k_{2z}}{\omega_2}\right)S_{j_1-1}^{l_1}\tilde{S}_{j_2-1}^{j_2+l_2-1}$$

$$-\left[p_{1z}p_{2\perp}q_z-p_{2\perp}(E_1+M)(E_q-M)\right]\left(1+\frac{\lambda_1 k_{1z}}{\omega_1}\right)\left(1-\frac{\lambda_2 k_{2z}}{\omega_2}\right)S_{j_1-1}^{l_1}\tilde{S}_{j_2+1}^{j_2+l_2}$$

$$+\left[p_{1\perp}p_{2z}q_z-p_{1\perp}(E_2+M)(E_q+M)\right]\left(1-\frac{\lambda_1 k_{1z}}{\omega_1}\right)\left(1+\frac{\lambda_2 k_{2z}}{\omega_2}\right)S_{j_1+1}^{l_1+1}\tilde{S}_{j_2-1}^{j_2+l_2-1}$$

$$-p_{1\perp}p_{2z}q_\perp\frac{\lambda_1 k_{1\perp}}{\omega_1}\left(1+\frac{\lambda_2 k_{2z}}{\omega_2}\right)S_{j_1}^{l_1+1}\tilde{S}_{j_2-1}^{j_2+l_2-1}-p_{1z}p_{2\perp}q_\perp\frac{\lambda_1 k_{1\perp}}{\omega_1}\left(1-\frac{\lambda_2 k_{2z}}{\omega_2}\right)S_{j_1}^{l_1}\tilde{S}_{j_2+1}^{j_2+l_2}$$

$$+p_{1z}p_{2\perp}q_\perp\frac{\lambda_2 k_{2\perp}}{\omega_2}\left(1+\frac{\lambda_1 k_{1z}}{\omega_1}\right)S_{j_1-1}^{l_1}\tilde{S}_{j_2}^{j_2+l_2-1}+p_{1\perp}p_{2z}q_\perp\frac{\lambda_2 k_{2\perp}}{\omega_2}\left(1-\frac{\lambda_1 k_{1z}}{\omega_1}\right)S_{j_1+1}^{l_1+1}\tilde{S}_{j_2}^{j_2+l_2}$$

$$+ p_{1\perp} p_{2\perp} q_\perp \frac{\lambda_1 k_{1\perp} \lambda_2 k_{2\perp}}{\omega_1 \omega_2} S_{j_1}^{l_1+1} \tilde{S}_{j_2}^{j_2+l_2-1} - p_{1\perp} p_{2\perp} q_z \frac{\lambda_1 k_{1\perp}}{\omega_1} \left(1 - \frac{\lambda_2 k_{2z}}{\omega_2}\right) S_{j_1}^{l_1+1} \tilde{S}_{j_2+1}^{j_2+l_2}$$

$$- p_{1\perp} p_{2\perp} q_z \frac{\lambda_2 k_{2\perp}}{\omega_2} \left(1 - \frac{\lambda_1 k_{1z}}{\omega_1}\right) S_{j_1+1}^{l_1+1} \tilde{S}_{j_2}^{j_2+l_2-1} - p_{1\perp} p_{2\perp} q_\perp \left(1 - \frac{\lambda_1 k_{1z}}{\omega_1}\right)\left(1 - \frac{\lambda_2 k_{2z}}{\omega_2}\right) S_{j_1+1}^{l_1+1} \tilde{S}_{j_2+1}^{j_2+l_2} \Bigg\},$$

(A11)

$$\upsilon_{21} = -i \int \frac{dq_\perp q_\perp}{(q^2 - M^2)} \Big\{ \Big[ p_{1z} p_{2z} q_z - p_{1z}(E_2+M)(E_q+M) + p_{2z}(E_1+M)(E_q-M) $$

$$- q_z(E_1+M)(E_2+M) \Big] \frac{\lambda_1 k_{1\perp}}{\omega_1} \left(1 - \frac{\lambda_2 k_{2z}}{\omega_2}\right) S_{j_1}^{l_1} \tilde{S}_{j_2+1}^{j_2+l_2+1}$$

$$+ \Big[ p_{1z} p_{2z} q_z + p_{1z}(E_2+M)(E_q+M) - p_{2z}(E_1+M)(E_q-M)$$

$$- q_z(E_1+M)(E_2+M) \Big] \frac{\lambda_2 k_{2\perp}}{\omega_2} \left(1 - \frac{\lambda_1 k_{1z}}{\omega_1}\right) S_{j_1+1}^{l_1} \tilde{S}_{j_2}^{j_2+l_2}$$

$$- \Big[ p_{1z} p_{2z} q_\perp - q_\perp(E_1+M)(E_2+M) \Big] \frac{\lambda_1 k_{1\perp} \lambda_2 k_{2\perp}}{\omega_1 \omega_2} S_{j_1}^{l_1} \tilde{S}_{j_2}^{j_2+l_2}$$

$$+ \Big[ p_{1z} p_{2\perp} q_z + p_{2\perp}(E_1+M)(E_q-M) \Big] \frac{\lambda_1 k_{1\perp} \lambda_2 k_{2\perp}}{\omega_1 \omega_2} S_{j_1}^{l_1} \tilde{S}_{j_2}^{j_2+l_2+1}$$

$$- \Big[ p_{1\perp} p_{2z} q_z + p_{1\perp}(E_2+M)(E_q+M) \Big] \frac{\lambda_1 k_{1\perp} \lambda_2 k_{2\perp}}{\omega_1 \omega_2} S_{j_1}^{l_1-1} \tilde{S}_{j_2}^{j_2+l_2}$$

$$- \Big[ p_{1\perp} p_{2z} q_z - p_{1\perp}(E_2+M)(E_q+M) \Big] \left(1 + \frac{\lambda_1 k_{1z}}{\omega_1}\right)\left(1 - \frac{\lambda_2 k_{2z}}{\omega_2}\right) S_{j_1-1}^{l_1-1} \tilde{S}_{j_2+1}^{j_2+l_2+1}$$

$$+ \Big[ p_{1z} p_{2\perp} q_z - p_{2\perp}(E_1+M)(E_q-M) \Big] \left(1 - \frac{\lambda_1 k_{1z}}{\omega_1}\right)\left(1 + \frac{\lambda_2 k_{2z}}{\omega_2}\right) S_{j_1+1}^{l_1} \tilde{S}_{j_2-1}^{j_2+l_2}$$

$$+ \Big[ p_{1z} p_{2z} q_\perp - q_\perp(E_1+M)(E_2+M) \Big] \left(1 - \frac{\lambda_1 k_{1z}}{\omega_1}\right)\left(1 - \frac{\lambda_2 k_{2z}}{\omega_2}\right) S_{j_1+1}^{l_1} \tilde{S}_{j_2+1}^{j_2+l_2+1}$$

$$- p_{1z} p_{2\perp} q_\perp \frac{\lambda_1 k_{1\perp}}{\omega_1} \left(1 + \frac{\lambda_2 k_{2z}}{\omega_2}\right) S_{j_1}^{l_1} \tilde{S}_{j_2-1}^{j_2+l_2} - p_{1\perp} p_{2z} q_\perp \frac{\lambda_1 k_{1\perp}}{\omega_1} \left(1 - \frac{\lambda_2 k_{2z}}{\omega_2}\right) S_{j_1}^{l_1-1} \tilde{S}_{j_2+1}^{j_2+l_2+1}$$

$$+ p_{1\perp} p_{2z} q_\perp \frac{\lambda_2 k_{2\perp}}{\omega_2} \left(1 + \frac{\lambda_1 k_{1z}}{\omega_1}\right) S_{j_1-1}^{l_1-1} \tilde{S}_{j_2}^{j_2+l_2} + p_{1z} p_{2\perp} q_\perp \frac{\lambda_2 k_{2\perp}}{\omega_2} \left(1 - \frac{\lambda_1 k_{1z}}{\omega_1}\right) S_{j_1+1}^{l_1} \tilde{S}_{j_2}^{j_2+l_2+1}$$

$$- p_{1\perp} p_{2\perp} q_\perp \frac{\lambda_1 k_{1\perp} \lambda_2 k_{2\perp}}{\omega_1 \omega_2} S_{j_1}^{l_1-1} \tilde{S}_{j_2}^{j_2+l_2+1} - p_{1\perp} p_{2\perp} q_z \frac{\lambda_1 k_{1\perp}}{\omega_1} \left(1 + \frac{\lambda_2 k_{2z}}{\omega_2}\right) S_{j_1}^{l_1-1} \tilde{S}_{j_2-1}^{j_2+l_2}$$

$$- p_{1\perp} p_{2\perp} q_z \frac{\lambda_2 k_{2\perp}}{\omega_2} \left(1 + \frac{\lambda_1 k_{1z}}{\omega_1}\right) S_{j_1-1}^{l_1-1} \tilde{S}_{j_2}^{j_2+l_2+1}$$

$$+ p_{1\perp} p_{2\perp} q_\perp \left(1 + \frac{\lambda_1 k_{1z}}{\omega_1}\right)\left(1 + \frac{\lambda_2 k_{2z}}{\omega_2}\right) S_{j_1-1}^{l_1-1} \tilde{S}_{j_2-1}^{j_2+l_2} \Big\},$$

(A12)

$$\upsilon_{22} = -\int \frac{dq_\perp q_\perp}{(q^2 - M^2)} \Big\{ \Big[ p_{1z} p_{2z} q_z + p_{1z}(E_2 + M)(E_q + M) + p_{2z}(E_1 + M)(E_q - M)$$
$$+ q_z(E_1 + M)(E_2 + M) \Big] \frac{\lambda_1 k_{1\perp} \lambda_2 k_{2\perp}}{\omega_1 \omega_2} S_{j_1}^{l_1} \tilde{S}_{j_2}^{j_2+l_2} + \Big[ p_{1z} p_{2z} q_z - p_{1z}(E_2 + M)(E_q + M)$$
$$- p_{2z}(E_1 + M)(E_q - M) + q_z(E_1 + M)(E_2 + M) \Big] \left(1 - \frac{\lambda_1 k_{1z}}{\omega_1}\right)\left(1 + \frac{\lambda_2 k_{2z}}{\omega_2}\right) S_{j_1+1}^{l_1} \tilde{S}_{j_2-1}^{j_2+l_2-1}$$
$$- \Big[ p_{1z} p_{2z} q_\perp + q_\perp(E_1 + M)(E_2 + M) \Big] \frac{\lambda_1 k_{1\perp}}{\omega_1} \left(1 + \frac{\lambda_2 k_{2z}}{\omega_2}\right) S_{j_1}^{l_1} \tilde{S}_{j_2-1}^{j_2+l_2-1}$$
$$- \Big[ p_{1\perp} p_{2z} q_z - p_{1\perp}(E_2 + M)(E_q + M) \Big] \frac{\lambda_1 k_{1\perp}}{\omega_1} \left(1 + \frac{\lambda_2 k_{2z}}{\omega_2}\right) S_{j_1}^{l_1-1} \tilde{S}_{j_2-1}^{j_2+l_2-1}$$
$$- \Big[ p_{1z} p_{2\perp} q_z + p_{2\perp}(E_1 + M)(E_q - M) \Big] \frac{\lambda_1 k_{1\perp}}{\omega_1} \left(1 - \frac{\lambda_2 k_{2z}}{\omega_2}\right) S_{j_1}^{l_1} \tilde{S}_{j_2+1}^{j_2+l_2}$$
$$- \Big[ p_{1\perp} p_{2z} q_z + p_{1\perp}(E_2 + M)(E_q + M) \Big] \frac{\lambda_2 k_{2\perp}}{\omega_2} \left(1 + \frac{\lambda_1 k_{1z}}{\omega_1}\right) S_{j_1-1}^{l_1-1} \tilde{S}_{j_2}^{j_2+l_2}$$
$$+ \Big[ p_{1z} p_{2z} q_\perp + q_\perp(E_1 + M)(E_2 + M) \Big] \frac{\lambda_2 k_{2\perp}}{\omega_2} \left(1 - \frac{\lambda_1 k_{1z}}{\omega_1}\right) S_{j_1+1}^{l_1} \tilde{S}_{j_2}^{j_2+l_2}$$
$$- \Big[ p_{1z} p_{2\perp} q_z - p_{2\perp}(E_1 + M)(E_q - M) \Big] \frac{\lambda_2 k_{2\perp}}{\omega_2} \left(1 - \frac{\lambda_1 k_{1z}}{\omega_1}\right) S_{j_1+1}^{l_1} \tilde{S}_{j_2}^{j_2+l_2-1}$$
$$+ p_{1z} p_{2\perp} q_\perp \frac{\lambda_1 k_{1\perp} \lambda_2 k_{2\perp}}{\omega_1 \omega_2} S_{j_1}^{l_1} \tilde{S}_{j_2}^{j_2+l_2-1} - p_{1\perp} p_{2z} q_\perp \frac{\lambda_1 k_{1\perp} \lambda_2 k_{2\perp}}{\omega_1 \omega_2} S_{j_1}^{l_1-1} \tilde{S}_{j_2}^{j_2+l_2}$$
$$+ p_{1\perp} p_{2\perp} q_z \frac{\lambda_1 k_{1\perp} \lambda_2 k_{2\perp}}{\omega_1 \omega_2} S_{j_1}^{l_1-1} \tilde{S}_{j_2}^{j_2+l_2-1} + p_{1\perp} p_{2\perp} q_\perp \frac{\lambda_1 k_{1\perp}}{\omega_1} \left(1 - \frac{\lambda_2 k_{2z}}{\omega_2}\right) S_{j_1}^{l_1-1} \tilde{S}_{j_2+1}^{j_2+l_2}$$
$$- p_{1\perp} p_{2\perp} q_\perp \frac{\lambda_2 k_{2\perp}}{\omega_2} \left(1 + \frac{\lambda_1 k_{1z}}{\omega_1}\right) S_{j_1-1}^{l_1-1} \tilde{S}_{j_2}^{j_2+l_2-1}$$
$$+ p_{1\perp} p_{2z} q_\perp \left(1 + \frac{\lambda_1 k_{1z}}{\omega_1}\right)\left(1 + \frac{\lambda_2 k_{2z}}{\omega_2}\right) S_{j_1-1}^{l_1-1} \tilde{S}_{j_2-1}^{j_2+l_2-1}$$
$$+ p_{1\perp} p_{2\perp} q_z \left(1 + \frac{\lambda_1 k_{1z}}{\omega_1}\right)\left(1 - \frac{\lambda_2 k_{2z}}{\omega_2}\right) S_{j_1-1}^{l_1-1} \tilde{S}_{j_2+1}^{j_2+l_2}$$
$$- p_{1z} p_{2\perp} q_\perp \left(1 - \frac{\lambda_1 k_{1z}}{\omega_1}\right)\left(1 - \frac{\lambda_2 k_{2z}}{\omega_2}\right) S_{j_1+1}^{l_1} \tilde{S}_{j_2+1}^{j_2+l_2} \Big\}.$$

(A13)

Here the integral of the triple-Bessel product [28] is given by

$$S_n^m = \int_0^\infty dr\, r J_n(k_{1\perp} r) J_{m-n}(q_\perp r) J_m(p_{1\perp} r), \tag{A14}$$

$$\tilde{S}_n^m = \int_0^\infty dr\, r J_n(k_{2\perp} r) J_{m-n}(p_{2\perp} r) J_m(q_\perp r). \tag{A15}$$

**Acknowledgments**

This work is supported by the Ministry of Science and Technology of the People's Republic of China (Grant Nos. 2018YFA0404803 and 2016YFA0401102), the National Natural Science Foundation of China (Nos. 11875307, 11935008 and 11774415), the Strategic Priority Research Program of Chinese Academy of Sciences (Grant No. XDB16010000).